\documentclass[useAMS,usenatbib]{mn2e}
\usepackage{graphicx}

\title[Gas accretion onto spiral galaxies]
{Accretion of gas onto nearby spiral galaxies}

\author[F. Fraternali and J. J. Binney]
{F. Fraternali$^{1}$\thanks{E-mail:
filippo.fraternali@bo.astro.it}
and J. J. Binney$^{2}$\\
$^{1}$Department of Astronomy, University of Bologna, via Ranzani 1, 40127, Bologna, Italy\\
$^{2}$Rudolf Peierls Centre for Theoretical Physics, 1 Keble Road, Oxford, OX1 3NP, UK}

\newcommand {\hi} {{\rm H}\,{\small\rm I}}
\newcommand {\kms} {\,{\rm km\,s}^{-1}}
\newcommand {\pc} {\,{\rm pc}}
\newcommand {\kpc} {\,{\rm kpc}}

\newcommand {\mo}{\,{M}_\odot}
\newcommand{\yr}{\,{\rm yr}}
\newcommand{\Myr}{\,{\rm Myr}}
\newcommand{\Gyr}{\,{\rm Gyr}}
\newcommand{\K}{\,{\rm K}}
\newcommand {\moyr}{\,{M_\odot\,\rm yr}^{-1}}
\newcommand{\gsim}{\lower.7ex\hbox{$\;\stackrel{\textstyle>}{\sim}\;$}}
\newcommand{\lsim}{\lower.7ex\hbox{$\;\stackrel{\textstyle<}{\sim}\;$}}

\begin{document}

\date{Accepted xxx. Received xxx}

\pagerange{\pageref{firstpage}--\pageref{lastpage}} \pubyear{xxxx}

\maketitle

\label{firstpage}

\begin{abstract}

We present evidence for cosmological gas accretion onto spiral galaxies in
the local universe.  The accretion is seen through its effects on the
dynamics of the extra-planar neutral gas.  The accretion rates that we
estimate for two nearby spiral galaxies are of the order of their star
formation rates.  Our model shows that most of the extra-planar gas is
produced by supernova feedback (galactic fountain) and only $10-20$\% comes
from accretion.  The accreting material must have low specific angular
momentum about the disc's spin axis, although the magnitude of the specific
angular-momentum vector can be higher.  We also explore the effects of a hot
corona on the dynamics of the extra-planar gas and find that it is unlikely
to be responsible for the observed kinematical pattern and the source of
accreted gas.  However, the interaction with the fountain flow should
profoundly affect the hydrodynamics of the corona.

\end{abstract}


\section{Introduction}

Observations of several spiral galaxies have revealed 
thick \hi\ layers or \hi\ haloes surrounding the galactic discs 
\citep[e.g.][]{swa97,fra02,mat03}.
The kinematics of this extra-planar gas is characterized by:
(i) a decrease in rotation velocity $v_{\phi}$ in the vertical
   direction \citep{swa97, fra05};
(ii) vertical motions from and towards the disc \citep{boo07};
(iii) a general radial inflow \citep{fra01}.
In a previous paper \citep[][hereafter FB06]{fb06}  we developed the
galactic fountain model  of extra-planar \hi\ in which gas clouds
are shot upwards from regions of star formation in the disc, and then follow
ballistic trajectories before returning to the disc \citep[see also][]{col02}.
Comparing the predictions of this model with data from deep \hi\ surveys of
the edge-on galaxy NGC\,891 and the moderately inclined galaxy NGC\,2403 we
concluded that the model fails in two respects: 
(i) it under-predicts the difference
between the rotation rates of disc and extra-planar gas, and (ii) it predicts
that the extra-planar gas should on average be flowing out rather than in.
We showed that these failures could not be rectified by assuming that the
clouds were visible in the 21-cm line on only part of their trajectories. 

In this paper we extend the fountain model to include interaction with
extragalactic gas. This gas is expected to take two forms: (i) the hot
corona of gas at the virial temperature, and (ii) cold, infalling clouds. 
Standard cosmological models of galaxy formation imply the existence of
both types of gas. In fact, arguments based on the cosmic background
radiation and primordial nucleosynthesis, combined with the measured
baryonic masses of galaxies imply that most of the baryons in the Universe
must lie in extragalactic gas \citep[e.g.][]{whi91,som06}. 
Limits on the background radiation
levels at frequencies from the X-ray to radio bands imply that most of the gas
must be at low densities and hot, $T\gsim10^6\K$.  From the existence of Ca
II absorption lines in the spectra of stars at high Galactic latitudes,
\cite{spitzer} inferred the existence of such gas around the Milky Way, and
by mapping the  soft X-ray emission from the massive galaxy NGC\,5746 
\cite{SommerLarsen} have established the existence of such gas
around at least some galaxies. \cite{BenjaminDanly} showed that
observed \hi\ clouds of fountain gas will experience appreciable drag as
they move through the coronal gas. The head-tail morphology of many
compact high-velocity clouds around the Milky Way \citep{Brunsetal01}
confirms this prediction. In this paper we add this drag to our earlier
dynamical model and show that although it cannot account for the observed
velocities of extra-planar \hi, it {must} have major implications for the
hydrodynamics of the coronal gas.

The existence of streams of cold gas around galaxies is less securely
established, but is supported by several lines of argument. From a
theoretical perspective, extended Press-Schechter theory
\citep{Bond,LaceyCole} predicts that galaxies grow through a series of infall
events. These range from a small number of major mergers, right down to an
almost continuous drizzle  of 
infalling dwarf galaxies and gas clouds.
It is
hard to assess the importance of this drizzle from numerical simulations of
limited mass and spatial resolution, but as simulations have become more
sophisticated, awareness of the importance of the drizzle has increased
\citep{birnb03,Keres,SemelinCombes,DekelB06}.

The rate of star-formation in the solar neighborhood has been remarkably
constant over the Galaxy's life \citep{twarog,BinneyDB}, which suggests that
the gas density has not decreased significantly even though the current
star-formation rate would exhaust the gas in a couple of gigayears. These
facts strongly suggest that gas consumed by star formation is mostly
replaced by accretion, whether by cooling of the corona or cold infall.
Steady accretion of metal-poor gas would also explain the discrepancy
between the observed stellar metallicity distribution in the solar
neighbourhood and that predicted by the closed-box model of chemical
evolution \citep{Tinsley,matteucci}.

What remains unclear is whether gas is predominantly accreted from the hot
corona or comes by cold infall, and in the latter case, how important is
steady {drizzle} compared to the acquisition of gas in parcels associated with
discrete events such as the capture of the Saggitarius dwarf galaxy.  
In our Galaxy, a limit can be placed on the masses of infalling objects
because the solar neighbourhood, which is both old \citep{BinneyDB} and
cold, would be rapidly heated by the passage through it of objects with
masses $\gsim10^7\mo$ \citep{LaceyO,TothO}.  

An observational indication of the importance of cold infall is the fact
that the Galaxy is surrounded by a large number of clouds, the so-called
high-velocity clouds (HVCs) \citep{wak97} that can be observed in \hi\ and
on the average have negative radial velocities.  New distance estimates for
the large HVC complexes show that they are located in the upper halo of the
Milky Way and have masses of order 10$^{6-7}\mo$ \citep{wak07,wak08}.
Most HVCs are too far from the plane to be part of the Galactic fountain,
and those that have measured metallicities are too metal-poor to be
predominantly gas ejected from the star-forming disc \citep{tri03}. The net
infall rate $\dot M$ associated with high-velocity clouds seems to be $\dot
M\sim 0.2 \mo\yr^{-1}$ \citep{wak07,pee07}, which is {about} an order of
magnitude lower than the SFR of the Milky Way.

There are observations of infalling gas complexes also in external galaxies
{\citep[for a review see ][]{san08}.} A well known example is the massive
\hi\ complex accreted by M\,101 \citep{vdh88} but the majority of these
accreting clouds will have much lower masses and have escaped detection.  In
fact, new deep observations of the region of space around M31 reveals the
presence of numerous HVCs with masses down to a few 10$^4\mo$ \citep{wes05}.
In some cases, halo clouds are detected at very anomalous ({\it
counter-rotating}) velocities with respect to the disc rotation and massive
cold filaments are also observed \citep{oos07,fra02}.  Gas accretion may be
also arise from interactions with satellite galaxies \citep{vdh05, san08}.

The above data suggest that star forming galaxies are accreting
material from the intergalactic space at a rate of {at least
$\dot M\sim 0.2 \mo\yr^{-1}$ \citep{san08}}
which is typically $10-20$\% of the SFR for these objects.
In this paper we argue that this directly observed accretion rate is
in fact a lower limit.
Most of the accreting gas interacts with gas pushed up into the halo by
stellar feedback, and it is {\it observable} only
indirectly via the peculiar kinematics of the extra-planar gas.

\section{The model}

Our pure fountain model is described in \citet{fb06}.  We integrate orbits
of particles that are ejected from the disc of a spiral galaxy and
travel through the halo until they fall back to the disc.  The data require
that the particles are ejected nearly straight upwards. The kinetic energy
of ejection is $\lsim4$ percent of the mechanical energy produced by supernovae.
The potential of the galaxy comprises the contributions of four
components: stellar and gaseous discs, both described by exponential
profiles, a dark matter (DM) halo and a bulge, both described by double
power law profiles \citep{deh98}. The potential is constrained by the \hi\
rotation curve \citep{oos07, fra02}. After each timestep of the orbit
integrations, the positions and velocities of the particles are projected
along the line of sight to produce an artificial cube of positions
on the sky and radial velocities. 
This cube is then smoothed and compared with the observed \hi\ data cubes.
Typically, the mass and spatial resolutions of the simulations are two orders 
of magnitude higher than those of the data, therefore the global effect is that
of a smooth fountain flow (see also FB06).

\section{Interaction with the hot corona}

In this section we concentrate on the edge-on galaxy NGC\,891, for which
parameters of the corona can be obtained from X-ray observations
\citep{bre97,str04}.  We model the corona as a plasma of
temperature $T$ that is in equilibrium in the axisymmetric potential of the
galaxy $\Phi(R,z)$.  Then the hydrostatic equations in cylindrical
coordinates imply that the density is
 \begin{equation}
\rho(R,z)= A~ \exp \left[- \frac{\mu m_{\rm p}}{k T} 
\left(\Phi(R,z) - \frac{v_0^2}{2} \right) \right],
\label{eq_rho}
\end{equation}
 where $A$ is a constant, $\mu=0.61$ is the molecular weight of the plasma,
and $v_0$ is the speed at which the corona rotates (assumed to be
independent of radius).  We adopt the temperature $T=2.7 \times 10^6\,{\rm
K}$ determined from X-ray spectra by \cite{str04} and for several values of
$v_0$ determine $A$ by fitting the predicted X-ray luminosity $L_{\rm X}$
to that observed.  $L_{\rm X}$ was obtained by integrating the emission
predicted by equation (\ref{eq_rho}) over roughly the same regions as that
of the X-ray observations.  The emissivity was taken from the MEKAL recipe
\citep{mew86,lie95} that is embedded in the XSPEC package.

The X-ray distribution is not highly flattened, so the adopted values
of $v_0$ lie significantly below the local circular speed.

In reality a significant fraction of the observed X-ray luminosity will
arise from interfaces between cold clouds and the corona rather from the
main body of the corona.  Consequently, our procedure for determining $A$
will tend to overestimate the corona's density. This fact will reinforce the
conclusion reached below, that the corona cannot absorb  angular momentum
fast enough to account for the observed rotational lag of the \hi\ halo.

\subsection{Cloud dynamics}

We consider the motion of an \hi\ cloud with temperature $\lsim10^4\K$
through the corona.  
Near the disc, the corona's cooling time $t_{\rm cool}=m_{\rm
p}k T/(\Lambda \rho)$ is $\gsim100\Myr$, which is much longer than the flow
time 
\begin{equation}
t_{\rm flow}=D/v\sim1\Myr
\end{equation}
 around a cloud of diameter $D\sim100\pc$ 
at speed
$v\sim100\kms$, so most of the coronal gas is expected to pass right by, and we should
idealize the interaction as the flow of a gas around a dense object. The
densities of the cloud and the corona differ by a factor $\gsim300$, so the
motion is akin to the flight through air of water drops from a garden
sprinkler. 

We assume that the flow of the corona past a cloud has a high Reynolds
number; this assumption minimizes the drag on the cloud, and we shall see
that even this minimum drag is highly significant.  When the Reynolds number
is high, the drag is proportional to the square of the flow velocity $v$ of
the corona past the cloud, so the acceleration of the cloud can be written
 \begin{equation}
\vec{g}_{\rm D}(R,z) =-{Cv\over L}\vec{v},
\label{eq:drag}
\end{equation}
 where $C$ is a dimensionless constant of order unity,
and $L$ is the distance over which the cloud interacts with its own mass of
coronal gas:
 \begin{equation}
L={m\over\rho\sigma},
\label{eq:defL}
\end{equation}
 with $\sigma$ and $m$ the cloud's cross-sectional area 
and mass,  respectively. In the absence of other forces, drag causes the cloud's
velocity to decay as
 \begin{equation}
\vec{v}(t)={\vec{v}_0\over1+Ctv_0/L}.
\end{equation}
 Thus the speed halves in a time $t_{\rm drag}=L/(Cv_0)\simeq300t_{\rm flow}$.
For $v_0\sim100\kms$, $t_{\rm drag}\sim300\Myr$ is of the same order as 
the orbital times 
of the clouds that make up the \hi\ halo (FB06 Fig.\ 10).
For low Reynolds numbers, $t_{\rm drag}$ would be even lower than the above
estimate. Note that if a cloud were to sweep up all the coronal gas that it
encounters rather than deflecting it, the drag on a cloud would be still given by
equations (\ref{eq:drag}) and (\ref{eq:defL}).

\subsection{Coronal spinup}

Since these numbers establish that momentum transfer between clouds and
coronal gas  may be
a significant process,  consider the possibility that this
process produces the additional lag of the \hi\ halo relative to the disc
that FB06 found the observations to require. Let fountain gas be ejected at
a rate $\dot M_{\rm f}$ and let $\delta\ell$ be the amount by which its
specific angular momentum is reduced by interaction with 
the corona. Then the rate at which the corona absorbs angular momentum is $\dot
L=\dot M_{\rm f}\delta\ell$, and the time required for the corona to be spun
up to near the circular speed is 
 \begin{equation}
t_{\rm spin}={M_{\rm corona}\ell_{\rm c}\over \dot M_{\rm f}\,\delta\ell},
\end{equation}
 where $\ell_{\rm c}=Rv_{\rm c}$ is the specific angular momentum of a
circular orbit.

Considering just the \hi\ that is more than $1.3\kpc$ from the plane, FB06
found that $\delta\ell/\ell_{\rm c}\sim0.05$ and $\dot M_{\rm f}=10^7\mo\Myr^{-1}$,
so $t_{\rm spin}\sim(2M_{\rm corona}/10^6\mo)\Myr$.  The mass of the corona
inside radius $r$ is $3\times10^7(r/10\kpc)\mo$, so $t_{\rm
spin}\sim60(r/10\kpc)\Myr$. That is, the capacity of the corona for
absorbing angular momentum is so slight that it will be brought to
corotation with the \hi\ halo within a dynamical time unless the angular
momentum is moved from the radii $\sim10\kpc$ at which it is imparted to
much larger radii. This cannot be done in less than a dynamical time. 

To test this conclusion, we added the deceleration (\ref{eq:drag}) to the
fountain model presented in FB06 -- a description of the initial parameters
of the model can be found in FB06.  Here the potential is taken to be the
one appropriate to a maximum disc, but in FB06 we showed that the kinematics
of the extra-planar gas is insensitive to the choice of potential. As
determined in FB06, the total mass of the \hi\ halo is $M_{\rm halo}=2
\times 10^9 \mo$.

We adjusted the ratio $m/\sigma$ until the model reproduced
the observed lag velocity.  We considered hot coronas rotating at different
speeds and with different velocity patterns.  In the case of a corona that
rotates at $100\kms$ independent of radius, the data for NGC\,891 are
successfully simulated when ${m}/{\sigma} \approx
10^5\mo\kpc^{-2}$ \citep[see also][]{fra07}.

\begin{figure}
\begin{center}
\includegraphics[width=240pt]{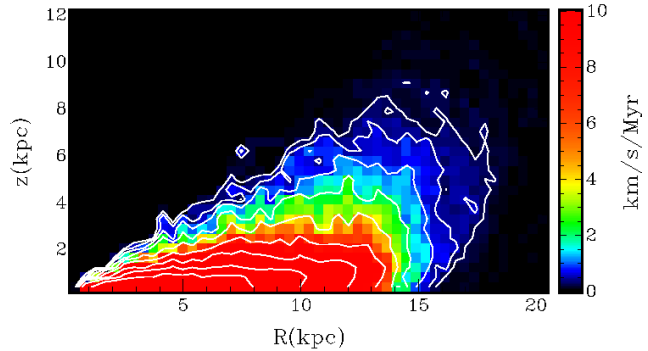}
\caption{Azimuthal velocities acquired by the hot corona per Myr
due to the transferring of angular momentum from the fountain gas.
From outside in contours are at 0.25, 0.5, 1, 2, 4, 8, 16, 32, 64 $\kms$/Myr.
\label{f_spinup}}
\end{center}
\end{figure}

However, in this picture, in which the coronal gas simply flows around clouds
without being accreted onto them, all the angular momentum lost by the \hi\
clouds of the halo has to be absorbed by the corona. If angular momentum
taken up by the corona in an annulus centred on $(R,z)$ remains in that
annulus, then in $1\Myr$ the azimuthal velocity of the corona at $(R,z)$
will increase by the amounts plotted in Fig.~\ref{f_spinup}; we see that the
spin-up time of the inner corona is extremely short; within a time $t_{\rm
spin} \approx\,10\Myr$ the inner corona will co-rotate with the disc and the
drag will vanish.

Thus the corona can be responsible for the observed rotational lag of the
\hi\ halo only if the clouds sweep up the coronal gas along their path
rather than deflecting it, which is not expected to happen given the large
ratio of the cooling time of coronal gas to the flow time around a cloud.
If clouds do not sweep up coronal gas, and the ratio $m/\sigma$ is not much
larger than the above estimate, the \hi\ halo will stir the corona rather
like the impeller of a centrifugal pump stirs the fluid in the pump.  In
response to this stirring, the coronal gas will move rapidly outwards
parallel to the disc plane, and be replaced by gas that moves down parallel
to the symmetry axis. Thus the \hi\ halo must drive a large-scale
circulation in the corona. Calculation of the dynamics of this circulation
lies beyond the scope of this paper, but is a topic that needs to be pursued
because it {must} have significant implications for the corona and X-ray
observations.

\section{Sweeping up of ambient gas}

In light of the conclusion above that the hot corona can be responsible for
the observed rotational lag only if it is swept up by fountain clouds, we
now examine the case in which fountain clouds sweep up ambient gas. Given
the long cooling time of the coronal gas, we envisage that the bulk of the
gas swept up is cold, infalling material.  When a fountain cloud impacts on
a stream of cold gas, some material is likely to be heated to the
virial temperature, but given the high densities involved, and the potential
for rapid turbulent mixing of the two bodies of gas, other material will be
swept up into the cloud of fountain gas. For simplicity we assume that this
is the dominant process. That is, we assume that a travelling cloud of mass
$m$ gains mass at a rate $\dot m=\alpha m$ and momentum at a rate $\dot
m\vec{v}_{\rm i}$, where $\vec{v}_{\rm i}$ is the velocity of the infalling
material.  We implement these assumptions by modifying the procedure used in
FB06 to determine the trajectories of clouds as follows. After each timestep
of duration $\delta t$ we increment the cloud's mass by $\delta m=\alpha
m\delta t$ and change its velocity from $\vec{v}_0$ to
 \begin{equation}
\vec{v}_1={m_0\vec{v}_0+\delta m\vec{v}_{\rm i}\over m_0+\delta m},
\end{equation}
 where $\vec{v}_{\rm i}$ is the velocity of the infalling material. 

The parameter $\alpha$, which is proportional to the volume density of
infalling material, becomes the key parameter to be constrained by the data.
Its spatial variation $\alpha(R,z)$ is not known.  However, experimenting
with different patterns we found that the results are not qualitatively very
different as long as its value does not vary too steeply in the inner halo. 
We therefore take
$\alpha$ to be constant, which implies that in a given interval $\delta t$ a
cloud captures the same amount of infalling gas regardless of its location.
We exclude regions very close to the disc, $|z|<0.2\kpc$.

A more critical characteristic of infalling material is the velocity
$\vec{v}_{\rm i}$ at which it is captured. At any
given time infalling material is likely to have some non-vanishing average
angular momentum, but the direction of this average changes over time
\citep{BinneyQuinn,Pichon}, and we neglect the correlation between its
average value at the present time and the angular-momentum vector of the
disc.

We consider two extreme possibilities: one possibility is that the $z$
component of $\vec{v}_{\rm i}$ is dominant and negative, implying that
infall is predominantly parallel to the $z$ axis.  The modulus of $v_{{\rm
i} z}$ is chosen at random between zero and the local escape speed.  The
other two components of $\vec{v}_{\rm i}$ are taken to follow Gaussian
distributions with zero mean and dispersion $\sim50\kms$.  The magnitude of
$\vec{v}_{\rm i}$ is constrained to be less than the escape speed.  With
this prescription, the angular momentum about the $z$ axis of an infalling
cloud vanishes in the mean.

The second possibility considered is that the dominant component of
$\vec{v}_{\rm i}$ is parallel to the local radius vector. For this component
we use the distribution that was previously used for $v_{{\rm i}z}$.  The
two tangential components of $\vec{v}_{\rm i}$ are drawn from a Gaussian
distribution with dispersion $\sim50\kms$, again subject to the constraint
that the total speed should not exceed the escape speed.  This prescription
may be motivated by the observation that in cosmological simulations
infalling material is typically on highly eccentric orbits \citep{Knebe04}.
On such an orbit, the radial component of the velocity vector is usually
dominant.  We also considered models with random velocity components
and no preferred directions; these gave results in between the two
extreme possibilities described here.

\subsection{Application to NGC\,891}

\begin{figure*}
\begin{center}
\includegraphics[width=400pt]{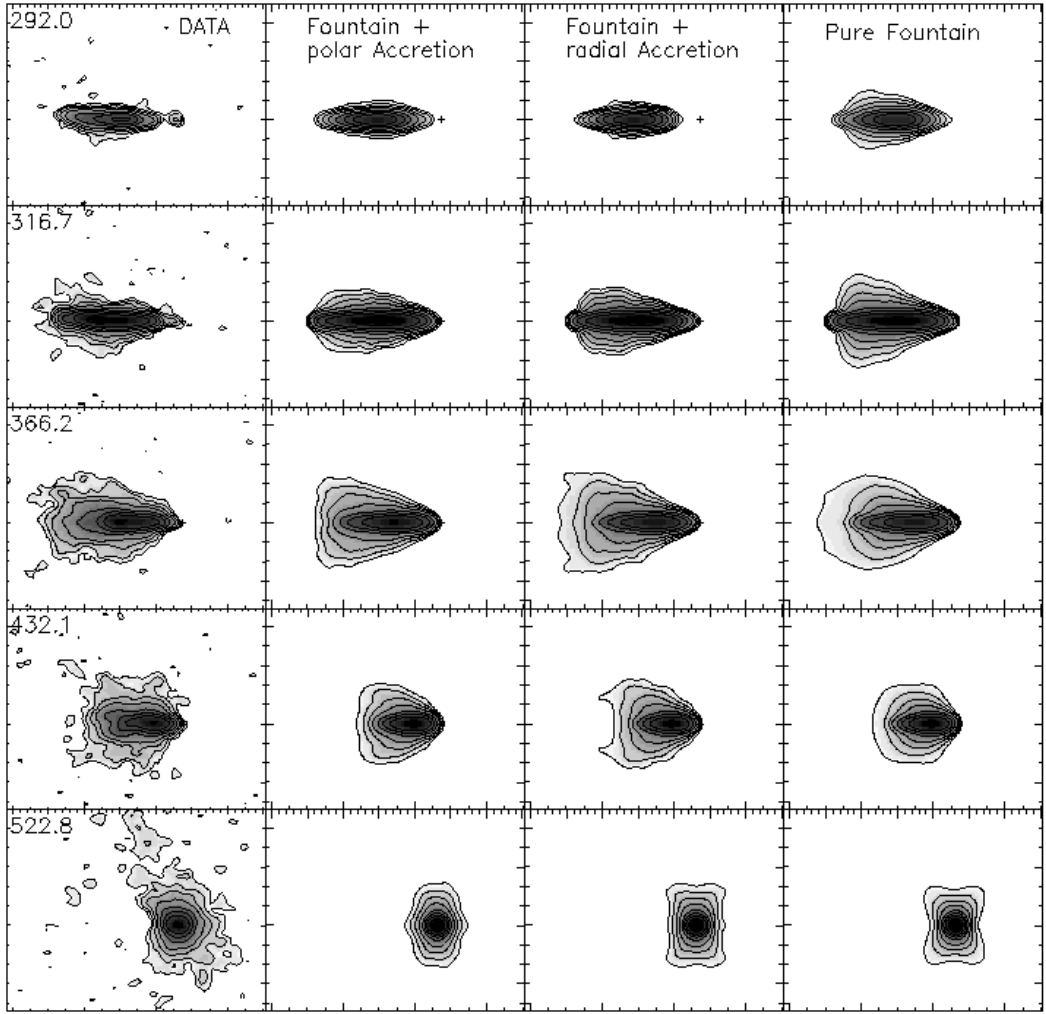}
\caption{
Comparison between 5 channel maps for NGC\,891 \citep{oos07}
and those obtained with three dynamical models. 
The first column shows the data, heliocentric radial velocities (in $\kms$)
are reported in the upper left corners. 
The channel map in the bottom row is roughly at the systemic velocity 
($V_{\rm sys}=528 \kms$).
Contour levels (for data and models) are: 0.25 (2.5$\sigma$), 5, 1, 2,
5, 10, 20, 50 mJy beam$^{-1}$.
\label{f_models891}}
\end{center}
\end{figure*}

We applied our model of a fountain that interacts with infalling gas to the
\hi\ observations of NGC\,891 \citep{oos07}. As in FB06, clouds were ejected
nearly vertically from the disc with a velocity drawn from a Gaussian
distribution, and the dispersion of this distribution was adjusted to
optimize the agreement between the model and observational data cubes.

Fig.~\ref{f_models891} (leftmost column) shows five sample channel maps for
NGC\,891 at $28''$ (about $1.3\kpc$) resolution; heliocentric
velocities in $\!\kms$ are in the top-left corners.  A key feature of the
data is that the apparent vertical thickness of the channel maps increases
as the velocity moves away from values characteristic of fast rotation
towards the systemic velocity.  Thus in the $v_{\rm hel}=292\kms$ channel
the \hi\ appears thin, and in the $v_{\rm sys}=528 \kms$ channel it has a
very extended vertical distribution.  This behaviour reflects the lag of the
$\hi$ halo with respect to the disc and is the main constraint on our models.

In the rightmost column of the same Fig.~\ref{f_models891} we report the
prediction from a model in which the extra-planar gas is produced by stellar
feedback only (pure galactic fountain), presented
in FB06.  Note that, thanks to the availability of a new deeper data cube
\citep{oos07} a lower contour is shown than appears in FB06.
These upper model maps show the distribution of \hi\ to be much thicker than
is found in the observed channel maps on the extreme left. This mismatch
arises because in the model of FB06 the rotation velocity decreases too slowly with increasing $z$.
FB06 show that this  problem cannot be eliminated by using different initial
conditions or supposing that the gas is invisible in the first part
of its orbit (phase-change models).

\begin{table}
 \centering
 \begin{minipage}{140mm}
   \caption{Fountain $+$ accretion models for NGC\,891}
   \label{t_models891}
   \begin{tabular}{@{}lccccc@{}}
     \hline
     Model & Phase & $h_{\rm v}$ & $\alpha$ & $\dot{\rm M}_{\rm in}$ &$\frac{\dot{\rm M}_{\rm in}}{\dot{\rm M}_{\rm out}}$\\
     & change & ($\kms$) &  (Gyr$^{-1}$) & ($\moyr$) & \\
     \hline
     Polar  & no & 90 & 2.0 & 3.4 & 0.12 \\
     Radial & yes & 80 & 0.6 & 2.3 & 0.05 \\
     \hline
   \end{tabular}
 \end{minipage}
\end{table}

Consider now the models shown in the  two central columns of
Fig.~\ref{f_models891}.  In these models the clouds of the fountain interact
with an accretion flow as described above, and differ from each other in the
dominant component of the infall velocity $\vec{v}_{\rm i}$: in the left
column $v_{{\rm i}z}$ is dominant, while in the right column $v_{{\rm i}r}$
dominates.
The specific accretion rates are
$\alpha=2\Gyr^{-1}$ and $\alpha=0.6\Gyr^{-1}$ respectively for the two models
(see Table \ref{t_models891})

It is evident that the top two channel maps of the left-central column have
narrower vertical distributions than are obtained without infalling gas, and
thus provide much better fits to the data.  Modest improvements in the shape
of the lower two channel maps for velocities near systemic are also evident.
We find that the 
total accretion rate of cold gas required by the model with dominant 
polar component of accretion is $\dot M_{\rm
inflow} \simeq 3.4\moyr$, which is very close to the star formation
rate of NGC891 \citep[$\hbox{SFR}=3.8\moyr$;][]{pop04}.

When the dominant component of the infall velocity is radial, the data are
best fitted when the fountain gas suffers a phase change such that it
becomes visible as \hi\ when $|z|$ peaks.  A galaxy with the star-formation
rate of NGC\,891 produces a significant flux of ionizing radiation and
it is plausible that a large fraction of the outgoing fountain flow would be
ionised.  
The right-central column shows channel maps for a model in which
cold material is accreting at a rate $\dot M_{\rm inflow} \simeq 2.3\moyr$.
We see that this model fits the data as well as the model with
polar accretion shown in the left-central column.

The two accretion models presented here are best fits to the data obtained
via a minimization of the residuals between the data and the models, much as
described in FB06 except that here the discrepancies are also considered
channel by channel to the whole data cube and not only in the total $\hi$
maps.  Once the mass of the halo has been fixed ($M_{\rm
halo}=2\times10^9\mo$), the key parameters are the accretion parameter
$\alpha$ (related to the accretion rate $\dot M_{\rm inflow}$) and, as in
the pure fountain model, the characteristic kick velocity.  For the latter
we found values slightly higher than in the pure fountain model
(80$-90\kms$) because to reach a given height fountain gas has to overcome
drag from infalling gas in addition to the galaxy's gravitational field. The
total energy input from the disc required to produce such a fountain flow is
$\lsim 6\,$ ($\lsim 8\,$ for the phase change model) per cent of the energy
available from the supernovae.  These percentages are a factor $\sim1.5$
larger than those of the pure fountain model (FB06).  The actual value of
the supernova feedback in galaxies is uncertain; our values agree with the
prediction of hydrodynamical models \citep[e.g.][]{maclow99}. The inflowing
(accretion) is only $\sim10$ percent of the outflowing (fountain) gas (see
Table~\ref{t_models891}).

In the model with radial accretion and phase change the accretion rate is
not well constrained and the data allow for values of $\alpha$ between 0.2
and 1.5 $\Gyr^{-1}$.  This is not surprising since this phase-change model
produces a fairly good fit to the data of NGC891 also without accretion, as
shown in FB06.  However in the case of low accretion rate it fails
completely to reproduce the inflow pattern observed in NGC\,2403.  We show
in the next section that the inclusion of accretion elegantly solves this
problem.

In the above models we assumed that the angular momentum of the accreting
material about the disc's spin axis, $L_z$, is zero on average.  We have
tested whether this is a necessary requirement for the model and found that
the data allow for a non-negligible $L_z$.  At radius $R$ the material
accreting on a disc with circular speed $v_{\rm c}$ must have on average
$L_{\rm z} < \frac12Rv_{\rm c}$. If $L_{\rm z}=\frac12Rv_{\rm c}$, the
accretion rate required to reproduce the data becomes $\dot M_{\rm
inflow}\sim6 \moyr$.
We discuss the implications of this requirement 
in Section \ref{conclusions}.

\begin{figure}
\begin{center}
\includegraphics[width=240pt]{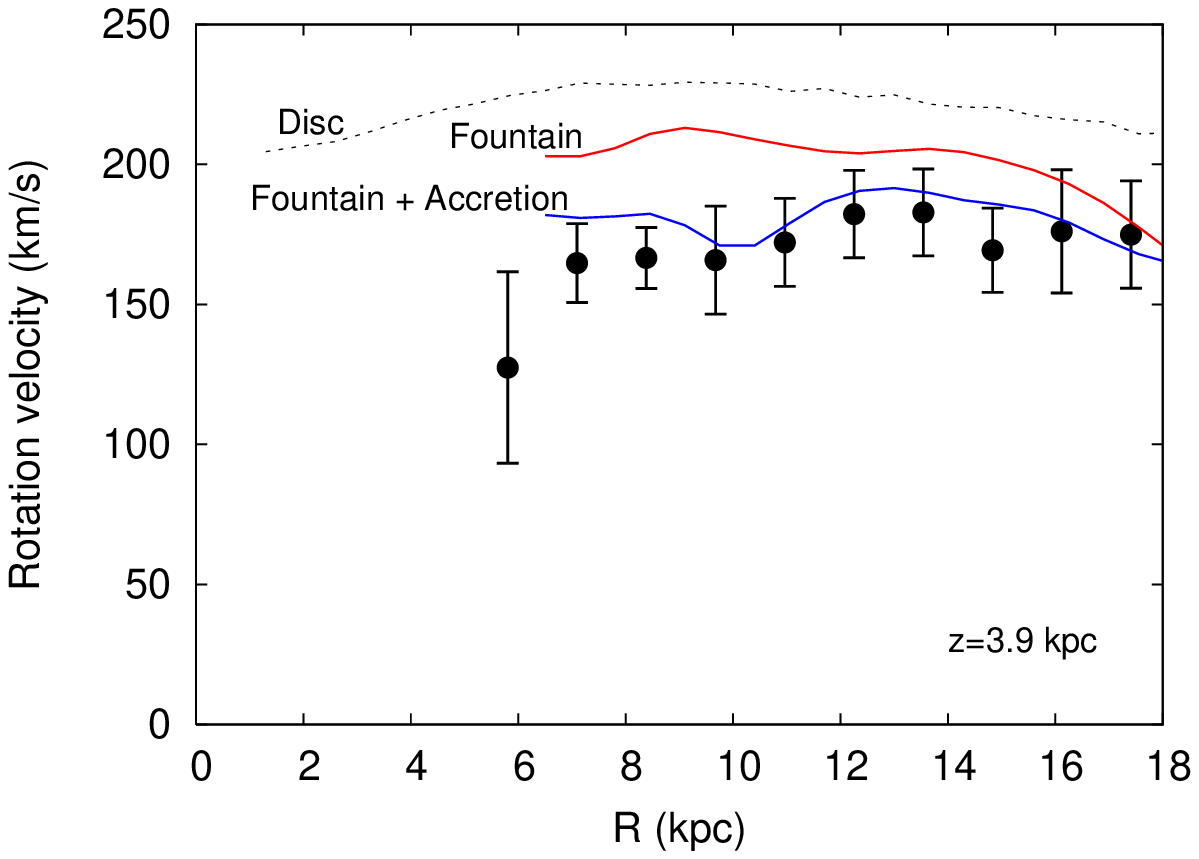}
\includegraphics[width=240pt]{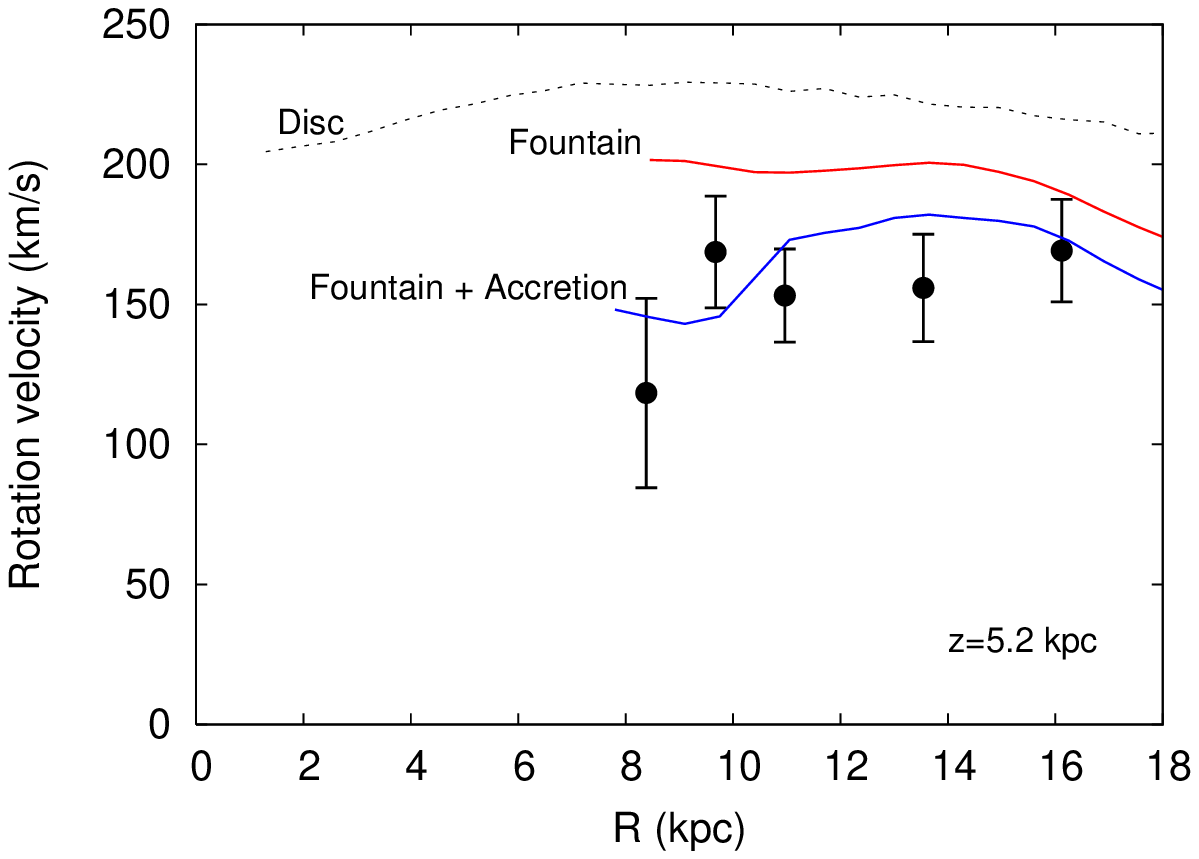}
\caption{
Rotational velocities (points) obtained from the \hi\ data 
at $3.9\kpc$ (upper panel) and $5.2\kpc$ (lower panel) 
from the plane of NGC\,891 \citep{fra05}.
The dotted line shows the rotation curve in the plane, while the other two 
lines show the azimuthal velocity predicted by our pure fountain 
(dotted-dashed/red) and fountain+accretion (solid/blue) models.
\label{f_rotcurves}}
\end{center}
\end{figure}

\cite{fra05} determined the rotation speed $v_{\rm rot}$
of gas in NGC\,891 at several heights above the plane by identifying the
terminal velocity $v_{\rm term}(R,z)$ at each position $(R,z)$. The rotation
speed is taken to be $v_{\rm rot}=v_{\rm term}-\sqrt{\sigma_{\rm
ran}^2+\sigma_{\rm obs}^2}$, where $\sigma_{\rm ran}$ is the random velocity
of \hi\ clouds and $\sigma_{\rm obs}=6\kms$ is the observational error.
\cite{fra05} assumed that $\sigma_{\rm ran} \simeq8\kms$ everywhere. More
recent modelling shows clear evidence for a larger dispersion $\sigma_{\rm
ran} \simeq20\kms$ away from the plane \citep{oos07}.  So it is appropriate
to move the points from \cite{fra05} down by $11\kms = \sqrt{20^2+6^2}-
\sqrt{8^2+6^2}$ (details in Fraternali, in prep.); the data points in Fig.~\ref{f_rotcurves} show the
corrected points for $z=3.9$ and $z=5.2\kpc$. The lines
show model predictions obtained from weighted means of the azimuthal
components of the particle velocities.  We see that the model with infall
(full/blue curves) fits the data points very well, while the pure fountain model
(dashed/red curves) yields only half the lag required by the data.

\subsection{Application to NGC\,2403}

\begin{figure}
\begin{center}
\includegraphics[width=240pt]{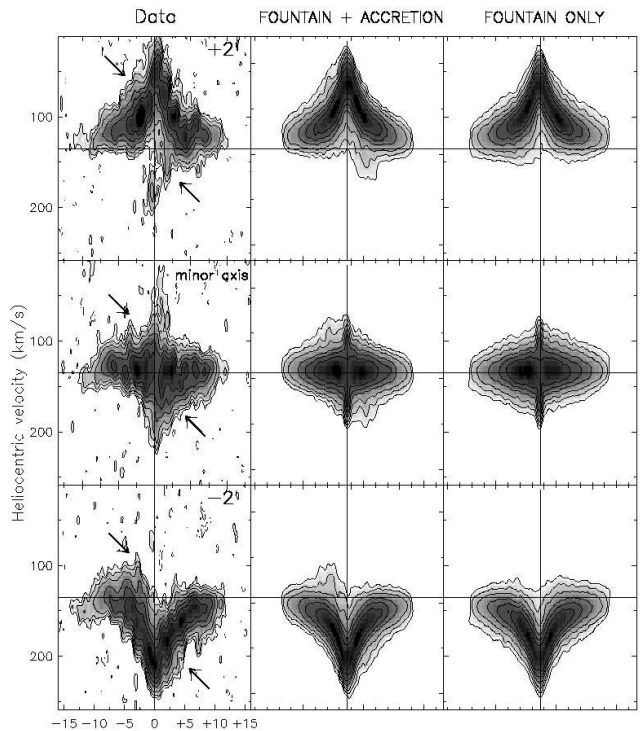}
\caption{
Comparison between 3 p-V plots parallel to the minor axis of NGC\,2403
and those obtained with two dynamical models. 
The first column shows the data,
contour levels (for data and models) are 0.5 ($\sim$2.5$\sigma$),
1, 2, 5, 10, 20, 50 mJy beam$^{-1}$.
\label{f_models2403}}
\end{center}
\end{figure}

Consider now the ability of the model that includes infall to reproduce the
data for the moderately inclined galaxy NGC\,2403.  \cite{fra02} showed that
the \hi\ measurements of this galaxy reveal an extended layer of
extra-planar gas that rotates less rapidly than the disc gas and flows in
towards the centre of the galaxy.  In FB06 we showed that pure fountain
models in which the halo gas rotates less fast than the disc show
outflow rather than inflow.
Here we show that adding accretion similar to that derived for NGC\,891 
causes the model to predict inflow similar to that observed in NGC\,2403.

Fig.~\ref{f_models2403} shows position-velocity plots along the minor axis
of NGC\,2403 (middle row) and parallel to this axis at offsets of $\pm2'$.  The leftmost
column shows the data, the rightmost column the pure fountain model and the
middle column the fountain with accretion.  Similar results are obtained
regardless of whether the accretion has a polar or radial pattern, and in
the latter case a phase change is not required (see Table \ref{t_models2403}).
The results shown are
obtained for the same values of the accretion parameter $\alpha$ as was
derived from NGC\,891
(for the radial infall model we doubled the value of $\alpha$ 
since clouds are now visible along the whole trajectory).
We see that the model with infall reproduces the data quite well. 
The required accretion rate is $\dot{\rm M}_{\rm inflow}\approx0.8\moyr$.
For comparison the SFR in NGC\,2403 is $1.2\moyr$ \citep{ken03}.

\begin{table}
 \centering
 \begin{minipage}{140mm}
   \caption{Fountain$+$accretion models for NGC 2403}
   \label{t_models2403}
   \begin{tabular}{@{}lccccc@{}}
     \hline
     Model & Phase & $h_{\rm v}$ & $\alpha$ & $\dot{\rm M}_{\rm inflow}$ &$\frac{\dot{\rm M}_{\rm in}}{\dot{\rm M}_{\rm out}}$\\
     & change & $(\!\kms)$ &  ($Gyr^{-1}$) & ($\moyr$) &\\
     \hline
     Polar & no & 55 & 2.0 & 1.0 & 0.21\\
     Radial & no & 50 & 1.2 & 0.6 & 0.15\\
     \hline
   \end{tabular}
 \end{minipage}
\end{table}

\begin{figure}
\begin{center}
  \includegraphics[width=200pt]{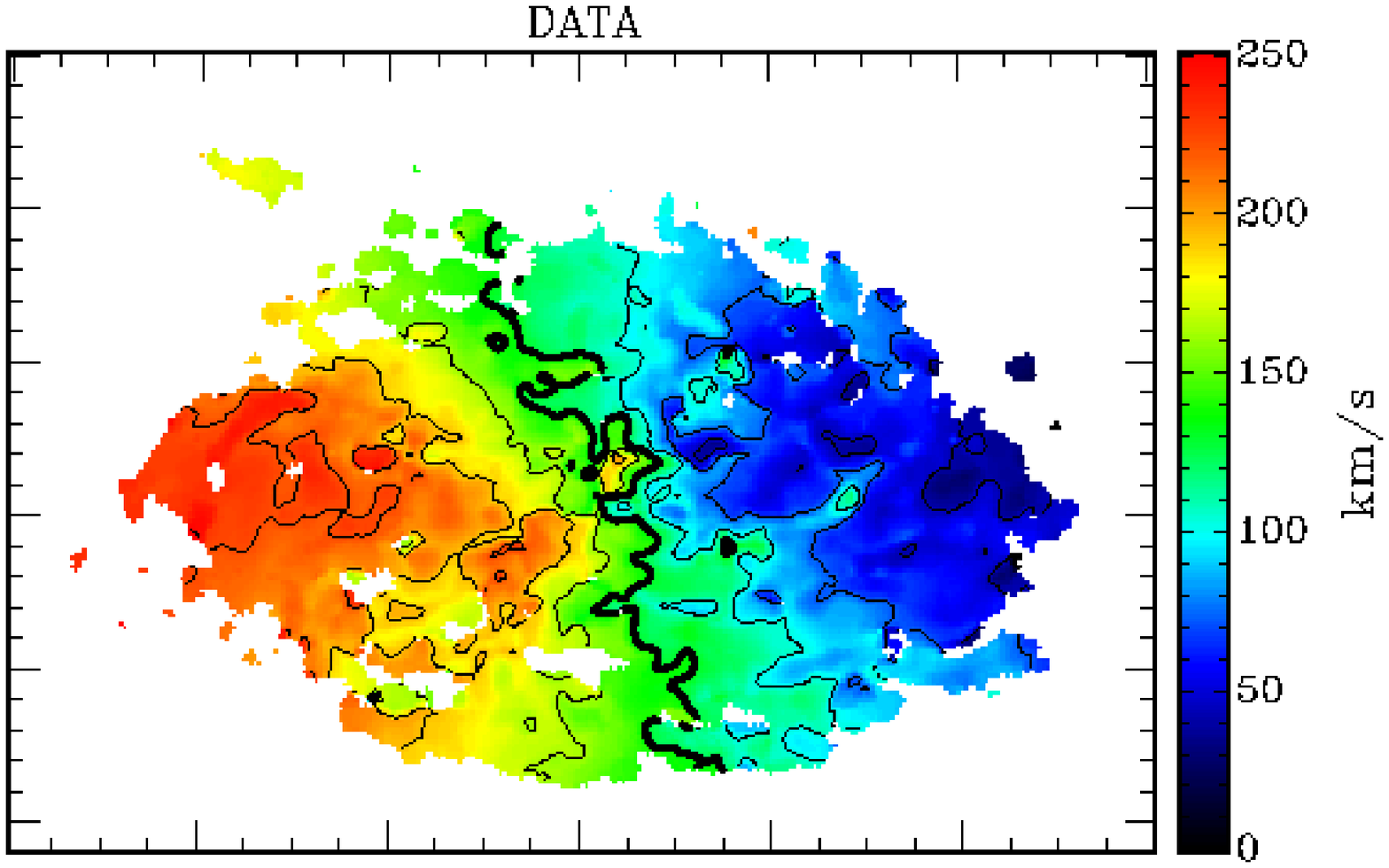}
  \includegraphics[width=200pt]{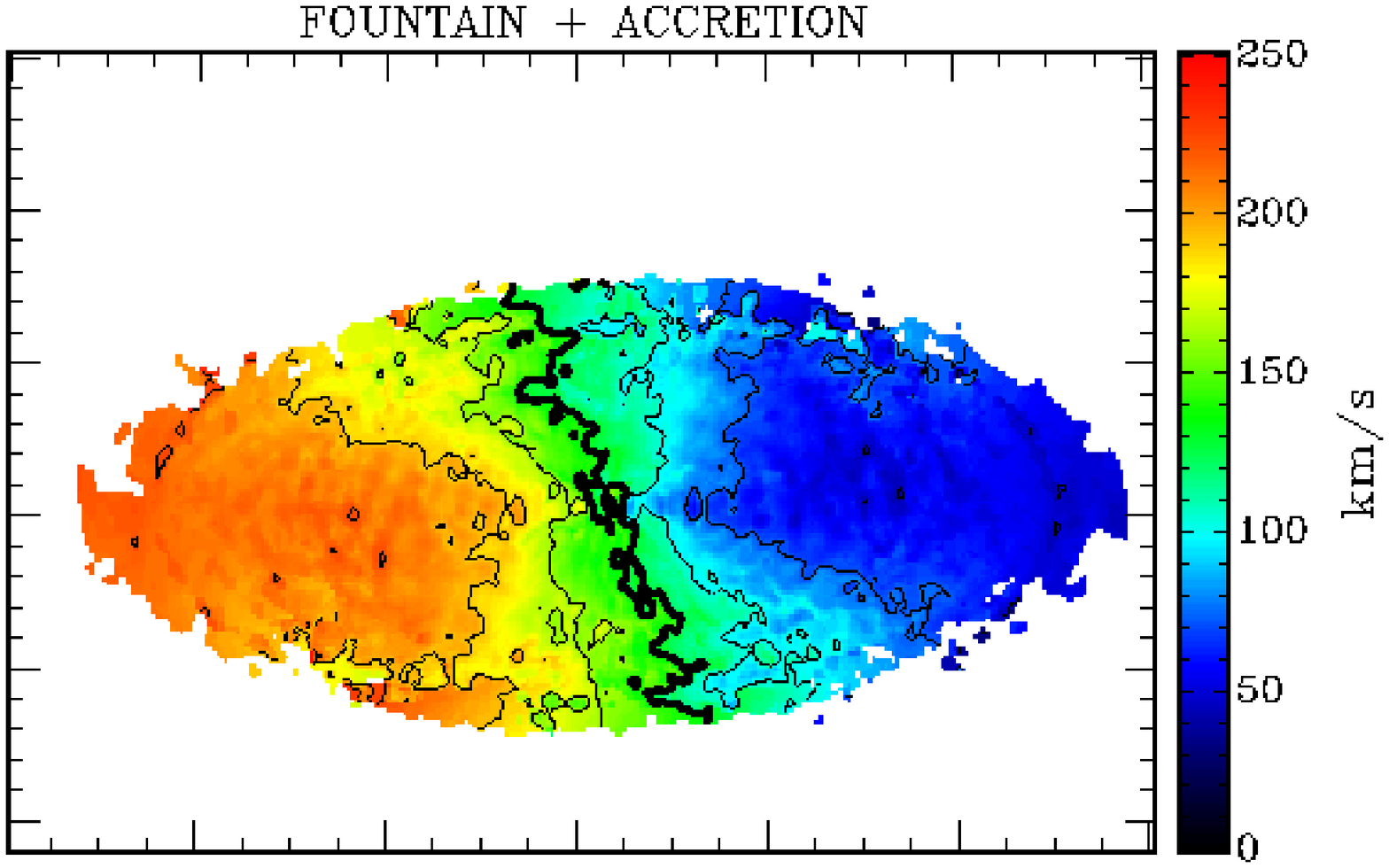}
  \includegraphics[width=200pt]{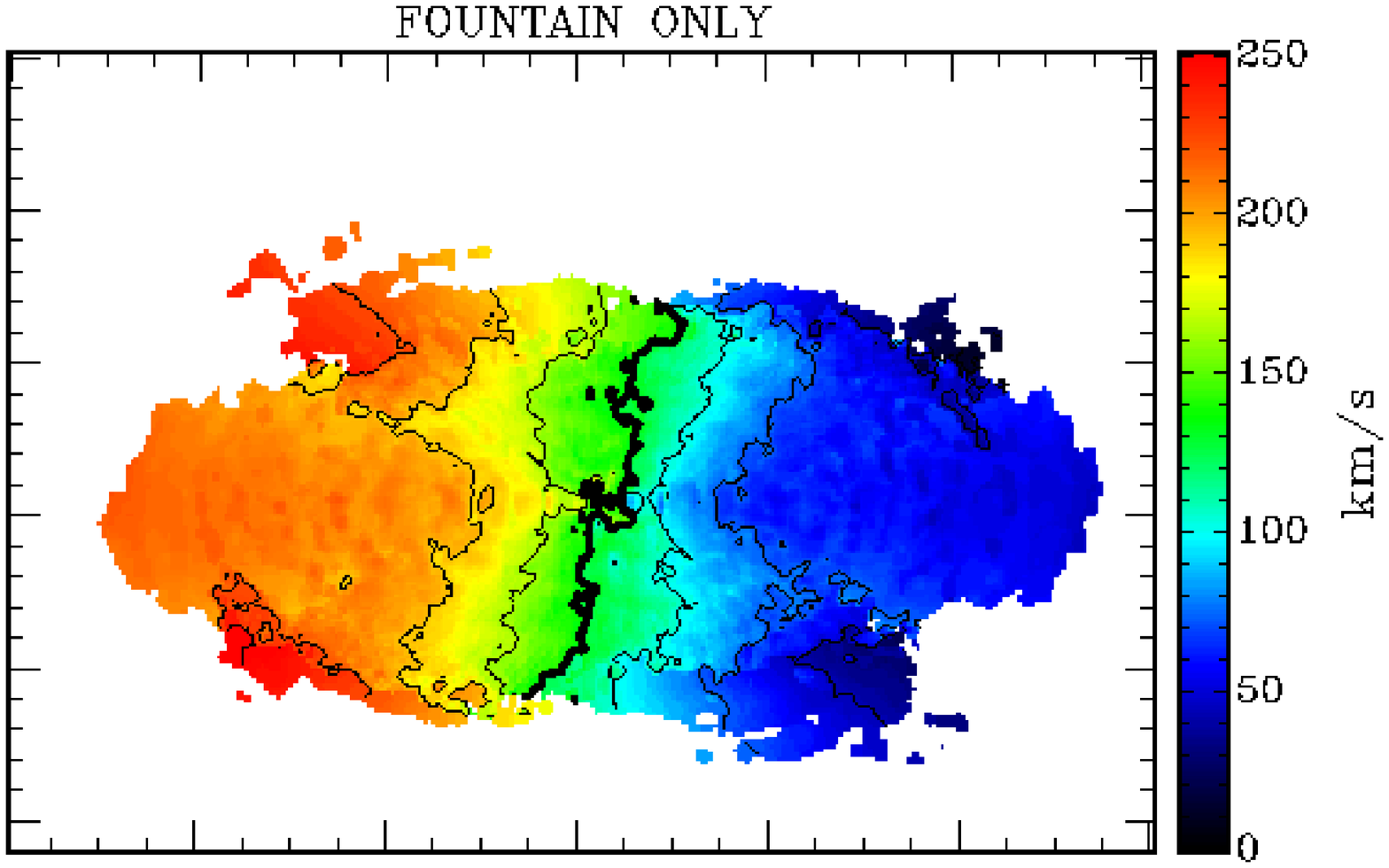}
  \caption{
Velocity field for the extra-planar (halo) gas of NGC\,2403 (upper panel) 
\citep[see][]{fra01}
compared with the two models
of fountain $+$ accretion (middle panel) and fountain only (bottom panel).
The velocity fields are rotated with the major axis of the galaxy along
the horizontal axis.
The tilt in the kinematic minor axis is the signature of inflow/outflow.
\label{f_velfi}
}
\end{center}
\end{figure}

\citet{fra01} have separated the so-called anomalous (extra-planar) gas 
in NGC\,2403 from the disc gas with a technique described in \citet{fra02}.
The resulting velocity field of the extra-planar gas 
(here shown in Fig.~\ref{f_velfi}, upper panel) shows a clear sign
of radial inflow in that the kinematical minor axis is rotated counter-clockwise
and not orthogonal to the major axis.
We have used the same procedure to derive velocity fields of the gas
above the plane in our model cubes.
The results are shown in the lower panels of Fig.\ \ref{f_velfi}.
Clearly the inclusion of gas accretion changes the direction of the clouds'
flow from outflow (bottom panel, pure fountain) to inflow (middle panel).
Note the tilt of the minor axis in the model is also quantitatively
very similar to the tilt seen in the data.

\begin{figure}
  \includegraphics[width=240pt]{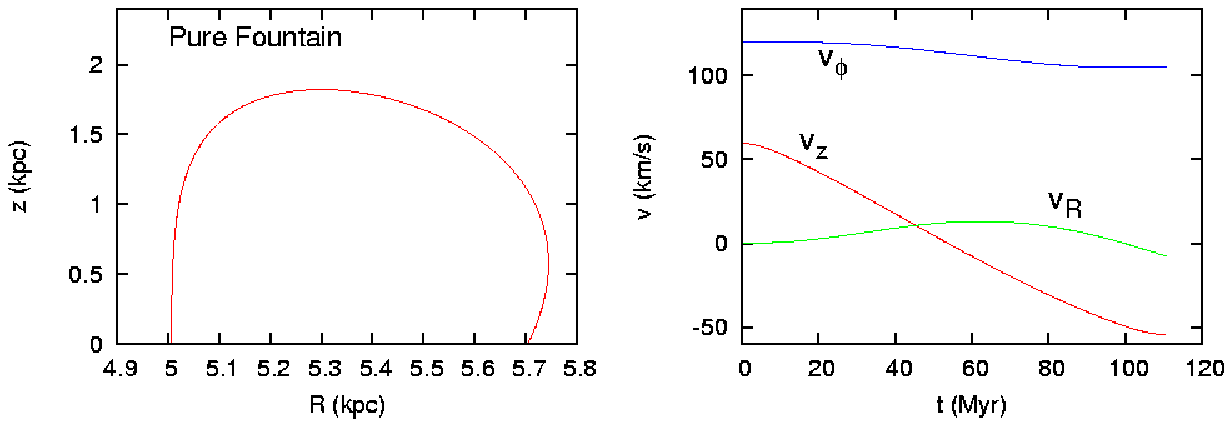}
  \includegraphics[width=240pt]{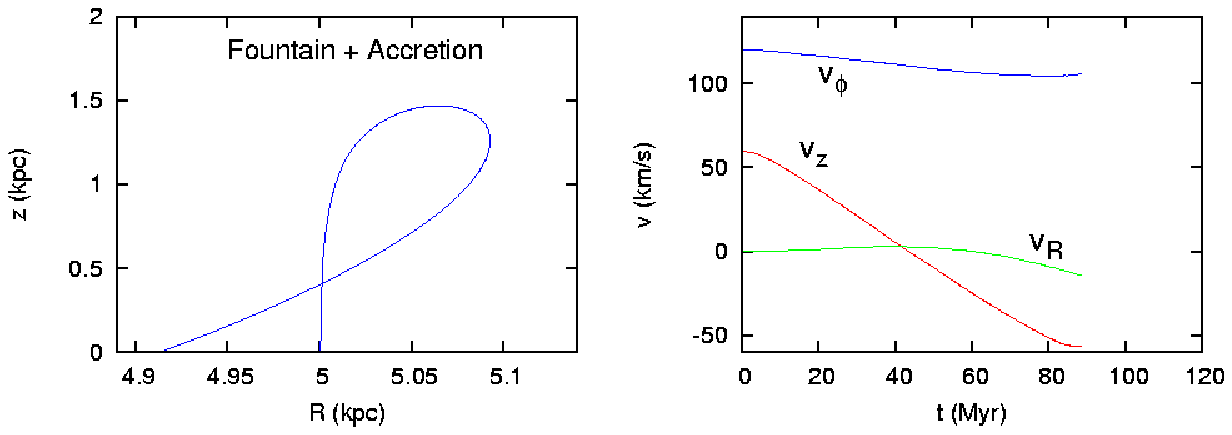}
  \includegraphics[width=240pt]{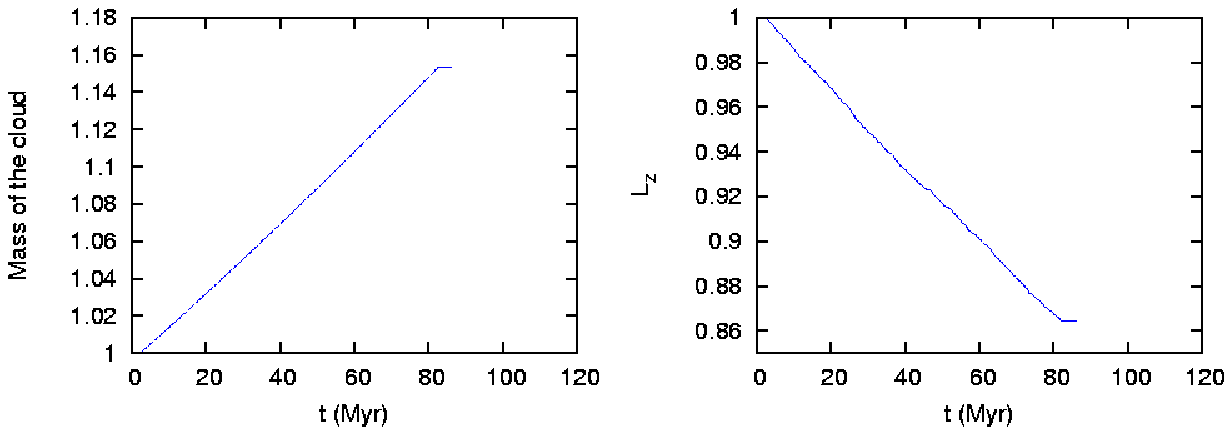}
  \caption{
    Trajectories of a representative particle in the potential of NGC\,2403.
The upper panels show the case of a pure fountain, the middle panels
show the effect of interaction with accreting material.
    In the bottom panel we show the modification of the mass and 
angular momentum of the fountain cloud as a consequence of this 
interaction.
\label{f_orbits}
}
\end{figure}

The above results occur because the orbits of fountain particles are
significantly changed by their interactions with infalling gas; 
in Fig.~\ref{f_orbits} the top left
and middle left panels show the orbit of a representative particle that is
launched at $\sim60\kms$ from $R\simeq5\kpc$ in the pure fountain model
(top) and the model with infall (middle; notice the different horizontal
scales in the two panels). In general the inclusion of accretion makes
fountain clouds fall back to the disc at radii that are roughly the same as
(or lower than)
the starting radii, rather than at larger radii as in classical fountain
models \citep[e.g.][]{col02}.
The bottom two panels of Fig.~\ref{f_orbits} show how
in the presence of infall the particle increases in mass and loses specific
angular momentum as a result of interactions.

The same accretion model (with the same accretion 
parameter) is able to produce both the
amount of lag of the extra-planar gas in NGC\,891 and the inflow pattern
required by NGC\,2403.
Qualitatively it is clear that when fountain particles interact
with the accreting material, they both have to share their angular momentum
and  acquire an inward
motion. However, it is remarkable that these effects are made quantitatively
correct by a single value  of $\alpha$.

\section{Relation to other work}

Our investigation of the coupling between clouds in the \hi\ halo and a
corona led to the conclusion that near the plane the corona would rapidly
approach corotation with the disc. Galaxies with rapidly spinning,
steady-state coronae were investigated by \citet{barnabe06}.  They found
that such coronae, taken in isolation, can reproduce the rotation curves
shown in Fig.~\ref{f_rotcurves}, raising the possibility that the observed
\hi\ clouds are embedded in and locally at rest with respect to a corona of
this type. Because the coronae described by Barnab\`e et al.\ enjoy
considerable rotational support, they are not very hot: in most of the
volume above the disc, $2\times10^4\K<T<2\times10^5\K$. At these
temperatures radiative cooling is at peak efficiency, and unless their
densities are made extremely low, the cooling times $\sim1\kpc$ above the
disc are much shorter than the local dynamical times. Similarly, if
densities are not very low, the coronae would be readily detectable in  UV
emission lines of ions such as O\,{\sc vi}.

A second model that has recently been used to fit the rotation curves above
the plane of NGC\,891 is based on cosmological inflow \citep{kau06}.  In
this model the gas is initially at the virial temperature and spinning. In a
nearly spherical region of radius $\sim10\kpc$ the gas rapidly cools and
collapses into the centre. Gas with low angular momentum moves in parallel
to the spin axis to take its place, causing the temperature to rise as one
approaches the axis at fixed $z$. Over several Gyr the remaining gas cools
onto an annular region of the disc. At this stage the cooling gas lags the
rotation of the disc in a manner similar to that observed in NGC\,891. In
fact, cooling from the initial conditions postulated by \cite{kau06} has
created a rapidly spinning corona similar to that discussed by
\citet{barnabe06}.

\cite{kau06} argue that the cooling corona will spawn clouds of
neutral \hi\ that constitute the observed \hi\ halo of NGC\,891. Their
simulations provide some support for this conjecture, and they argue that
the proposal is supported by the analytic work of \cite{MallerB}.  Since
the simulations are subject to numerical artifacts associated with their
finite resolution in mass and distance \citep{kau07}, the analytic
arguments for the existence of cool clouds are important.

We believe that these analytic arguments are flawed
because they ignore the ability of a
gravitational field to prevent a stratified atmosphere from suffering the
\cite{Field65} thermal instability
\citep{Malagoli_etal87,BalbusSoker89}.\footnote{In \cite{MallerB} the analysis
relies on \cite{Balbus86}. However in their equation (20) they misquote
Balbus's equation (5) by misidentifying ${\cal L}(T,\rho)$ as the
conventional cooling function $\Lambda(T)$.}
Physically, when an atmosphere
is confined by a gravitational field (which can be taken to include the
pseudo-field associated with spin), the gas settles on a dynamical time into
a stratified configuration, in which the material of lowest specific entropy
lies at the lowest potential. In this configuration an overdense region is a
region in which the surfaces of constant specific entropy have been
perturbed upwards. Consequently it is the crest of an internal gravity wave.
On a timescale that is only slightly longer than the dynamical time, the
overdensity sinks from lack of buoyancy, overshoots its point of
equilibrium, and is soon an underdense region.  Consequently the basis of
the Field instability -- the increase in cooling rate with increased density
-- cancels out to first order as initially overdense regions oscillate
vertically and are on average neither more nor less dense than undisturbed
material at their mean position \citep{BalbusSoker89}. 
There is no reason to
suppose that hot gas that is cooling onto the disc of a galaxy is more
liable to the Field instability than is the gas of cooling flows, which
X-ray spectra showed do not evolve according to the Field instability
\citep{PetersonEtal02}, as \cite{Malagoli_etal87} predicted more than a
decade earlier.

A second problem with the contention that \hi\ haloes represent material that
is condensing out of the corona is the massiveness of haloes. 
In NGC\,891, the \hi\ mass detected above 1 kpc from the plane is 
$1.2\times 10^9 \mo$, almost 30\% of the total \hi\ mass \citep{oos07}.
If this gas were falling towards the disk even at a relatively low speed of 
$\sim 100 \kms$, taking a mean distance from the plane of about 4 kpc, it 
would imply an accretion rate of $\sim30 \moyr$.
Not only does this rate substantially
exceed the current SFR, but it implies that we live in a time when the disc
is growing much faster than it has in the past, which is a priori
implausible and probably inconsistent with the measured colours of the
stellar disc.

Thirdly, \cite{RandBenjamin08} searched for \hi\ around the only spiral
galaxy that is known to have a large corona, the edge-on galaxy NGC\,5746.
Their finding that there is very little \hi\ in the corona is hard to
reconcile with the contention of \cite{kau06} that thermal instability of
coronae leads to the abundant formation of
\hi\ clouds.

{Finally, from an observational point of view there is strong evidence
that a large fraction of the extra-planar gas is indeed produced by the 
galactic fountain. 
The distribution of the extra-planar \hi\ in NGC\,891 and NGC\,2403 is 
concentrated very close to the star-forming disk and almost absent
above and below the outer disk.
In other galaxies like NGC\,6946 \citep{boo07} most of the 
high-velocity features are located in the inner star forming disc.
Moreover, \citet{hea07} found that in three galaxies with extra-planar 
ionised gas, the thickness of the ionised halo as well as the 
vertical rotational gradient both clearly correlate with the SFRs, pointing
to a dominance of the fountain in the formation of these haloes.
}

\section{Discussion and conclusions}
\label{conclusions}

Standard cosmology teaches that the majority of the baryons lie somewhere in
extragalactic space, probably in a hot diffuse medium.  We cannot hope to
have an adequate knowledge of the formation and evolution of galaxies until
we have a better understanding of the connection between star-forming
galaxies and extragalactic gas. In this paper we have shown that the
existence of \hi\ haloes around star-forming galaxies has profound
implications for this vital connection.

It has long been evident that star-forming galaxies must be accreting gas at
significant rates. We have modified the ballistic fountain model to include
the effects of clouds sweeping up material that 
has low angular momentum about the disc's spin axis. 
The
key parameter of the model is the rate $\alpha$ at which the mass of a cloud
exponentiates as a result of accretion. We find that a single value, $\alpha
\approx 1.5\Gyr^{-1}$ enables us to resolve the problems encountered in FB06
with both the edge-on galaxy NGC\,891 and the inclined galaxy NGC\,2403.
Specifically, adding swept-up gas doubles the lag in rotation of the halo with
respect to the disc because clouds have to share their angular momentum with
swept-up gas, and the any inward motion of accreted material causes the
net motion of the fountain gas to be inwards rather than outwards.  The
aggregate infall rates implied for NGC\,891 and NGC\,2403 are $\sim2.9$ and
$\sim0.8\moyr$, respectively. These rates are very close to the rates at
which star formation is consuming gas in the two galaxies, but  $1/5-1/10$
of the rates at which star-formation cycles gas through the halo (FB06).
This means that most of the neutral gas that we observe in the halo
($\sim$90\% in NGC\,891 and $\sim80-85$\% in NGC\,2403) is gas that stellar
feedback has pushed up from the disc and only a small fraction is accreted.
{Notably, 
the obtained accretion rates are not only consistent with the respective 
SFRs, but also with the expected cosmological accretion rates for these
types of galaxies \citep[e.g.][]{neist08}.}

It is unclear whether the swept-up gas arises from cooling of a hot medium,
or from the infall of cold streams of gas
\citep{birnb03,Bin04,Keres,Cattaneo06}, but the long cooling time of coronal
gas strongly suggests that it comes from cold streams.  The accretion rates
onto nearby spiral galaxies determined here are cosmologically significant:
if the two galaxies had accreted at their current rates for a Hubble time,
they would have added $\sim1/2-1/3$ of their current masses.  Hence our
estimates are consistent with past accretion rates being larger, but not
much larger unless SN feedback is efficient in ejecting material back to the
IGM.  Note that estimated accretion rates could be considered as lower
limits because gas can reach the discs without encountering fountain clouds,
either by slipping past these clouds if they have a small filling factor, or
by joining the disc at large radii, outside the star-forming region. 

The accretion rates estimated here exclude stars brought in by satellites,
but since accreted stars cannot contribute to the thin disc, the bulk of the
accretion onto these disc-dominated galaxies must be gaseous, consistent
with the significant rates of gas accretion that we require.

We have argued that feedback from star formation is essential for the
formation of \hi\ haloes -- the latter cannot form directly from cooling
coronal gas because (i) the latter is not thermally unstable, (ii) if \hi\
haloes represented material that has condensed from the corona, the
accretion rate of the thin disc would be implausibly large, and (iii) there
is broad observational evidence of a link between galactic fountains and
extra-planar gas features.

In the likely event that coronal gas is not swept up by fountain clouds, 
the corona must be profoundly influenced by interaction with fountain gas.
In particular, just above the disc interaction with halo clouds will ensure
that the corona is nearly corotating. If the nearly corotating gas is
confined near the disc, it must be quite cool ($T\sim10^5\K$) and strongly
radiating in the UV. It seems more likely that the corona is hotter and
therefore not bound to the vicinity of the disc: then it will flow
outwards and upwards. 

We have to expect that the interface between \hi\ in a galactic disc and
coronal gas is at least as complex and dynamic as that between the solar
photosphere and corona because similar physics is at work: global rotation,
local vortices, a steep density gradient, upswelling of material, magnetic
loops and optically-thin radiative cooling in the temperature range
$10^4-10^7\K$. It is worth noting that above the Sun the transition from
$T=3\times10^4\K$ to $T=3\times10^5\K$ is complete within only $\sim30\,$km
\citep{Jordan77}. 

Our assumption that swept-up gas has low angular momentum about the disc's
spin axis is both essential and at first sight surprising. Indeed, discs
must be formed by gaseous infall, and they are expected to form from inside
out as a result of the mean angular momentum of infalling material rising
over time.  Galaxies such as NGC\,4550, which has two counter-rotating
stellar discs \citep{Rubinetal} and the numerous decoupled gaseous/stellar
systems found with SAURON \citep{sarzi06} make it certain that whatever may
be true of the time-evolution of the magnitude of the specific
angular-momentum vector $\vec{L}$ of infalling gas, any given component
$L_z$ can diminish and even change sign. For our model it is essential only
that at the radius $R$ of accretion to a disc with circular speed $v_c$ we
have $L_z< {1\over2}Rv_{\rm c}$. The perpendicular components of $\vec{L}$
could well make  $|\vec{L}|>Rv_{\rm c}$. 
{In the future, it will be possible to test these predictions with
hydrodynamical cosmological simulations.
Existing models seem to agree qualitatively
with our requirements \citep[e.g.][]{porc02}.
}

{Moreover, one} interpretation of warps in \hi\ discs relies precisely on
$\vec{L}$ for infalling material being comparable in magnitude to, but
strongly misaligned with, $\vec{L}$ for the outer disc
\citep{OstrikerB,JiangB,ShenS}.  This argument indicates that there {could}
be a connection between disc warping and rotational lag of the \hi\ halo.
{Interestingly, both NGC\,891 and NGC\,2403 do show warping of the outer
discs \citep{ruiz02,fra02}. However, if the specific angular momentum of
infalling material is large enough, there could be negligible infall in the
vicinity of the fountain gas. Then the galaxy would have a warped \hi\ disc
but an \hi\ halo with only a small rotational lag.}

We have modelled the ambient gas as smooth, so  fountain clouds
receive a constant amount of mass and momentum per given time.  In reality,
an accretion flow is probably clumpy.  However,
as long as the gas mass of the accreting clouds is lower than the mass
resolution in the data (about $10^6 \mo$), no significant difference is
expected.  For larger accreted masses, the effects of accretion will be more
pronounced in some regions of the halo than in others, unless large gas
complexes break into smaller clumps, which can be modelled by smooth
accretion.

It would be interesting to extend this work to other galaxies, in particular
low surface-brightness (LSB) galaxies with low star formation rates that
also have \hi\ haloes -- UGC\,7321 \citep{uson03} is an example of such a
system.  This would allow us to establish if accretion is a common process
at the current epoch, and what role it plays in triggering star formation.
We are also applying the model to the Milky Way; this work will clarify the
relation between fountain gas and the classical phenomena of intermediate-
and high-velocity clouds (Fraternali \& Binney, in prep.).

\section*{Acknowledgments}

This work was in part  supported by Merton College, Oxford.

{}

\bsp

\label{lastpage}

\end{document}